\documentclass[%
 aip,
 amsmath,amssymb,
 reprint,%
]{revtex4-1}

\usepackage{graphicx}
\usepackage{dcolumn}
\usepackage{bm}

\usepackage[utf8]{inputenc}
\usepackage[T1]{fontenc}
\usepackage{mathptmx}
\usepackage{etoolbox}

\makeatletter
\def\@email#1#2{%
 \endgroup
 \patchcmd{\titleblock@produce}
  {\frontmatter@RRAPformat}
  {\frontmatter@RRAPformat{\produce@RRAP{*#1\href{mailto:#2}{#2}}}\frontmatter@RRAPformat}
  {}{}
}%
\makeatother
\begin{document}

\preprint{AIP/123-QED}
\title{Femtotesla Atomic Magnetometer for Zero- and Ultralow-field Nuclear Magnetic Resonance}
\author{Taizhou Hong}
\affiliation{
CAS Key Laboratory of Microscale Magnetic Resonance and School of Physical Sciences, University of Science and Technology of China, Hefei, Anhui 230026, China}
\affiliation{
CAS Center for Excellence in Quantum Information and Quantum Physics, University of Science and Technology of China, Hefei, Anhui 230026, China}
\affiliation{
\mbox{Hefei National Laboratory, University of Science and Technology of China, Hefei 230088, China}}

\author{Yuanhong Wang}
\affiliation{
CAS Key Laboratory of Microscale Magnetic Resonance and School of Physical Sciences, University of Science and Technology of China, Hefei, Anhui 230026, China}
\affiliation{
CAS Center for Excellence in Quantum Information and Quantum Physics, University of Science and Technology of China, Hefei, Anhui 230026, China}
\affiliation{
\mbox{Hefei National Laboratory, University of Science and Technology of China, Hefei 230088, China}}

\author{Zhenhan Shao}
\affiliation{
CAS Key Laboratory of Microscale Magnetic Resonance and School of Physical Sciences, University of Science and Technology of China, Hefei, Anhui 230026, China}
\affiliation{
CAS Center for Excellence in Quantum Information and Quantum Physics, University of Science and Technology of China, Hefei, Anhui 230026, China}
\affiliation{
\mbox{Hefei National Laboratory, University of Science and Technology of China, Hefei 230088, China}}

\author{Qing Li}
\affiliation{
CAS Key Laboratory of Microscale Magnetic Resonance and School of Physical Sciences, University of Science and Technology of China, Hefei, Anhui 230026, China}
\affiliation{
CAS Center for Excellence in Quantum Information and Quantum Physics, University of Science and Technology of China, Hefei, Anhui 230026, China}
\affiliation{
\mbox{Hefei National Laboratory, University of Science and Technology of China, Hefei 230088, China}}

\author{Min Jiang}
\email[]{dxjm@ustc.edu.cn}
\affiliation{
CAS Key Laboratory of Microscale Magnetic Resonance and School of Physical Sciences, University of Science and Technology of China, Hefei, Anhui 230026, China}
\affiliation{
CAS Center for Excellence in Quantum Information and Quantum Physics, University of Science and Technology of China, Hefei, Anhui 230026, China}
\affiliation{
\mbox{Hefei National Laboratory, University of Science and Technology of China, Hefei 230088, China}}

\author{Xinhua Peng}
\email[]{xhpeng@ustc.edu.cn}
\affiliation{
CAS Key Laboratory of Microscale Magnetic Resonance and School of Physical Sciences, University of Science and Technology of China, Hefei, Anhui 230026, China}
\affiliation{
CAS Center for Excellence in Quantum Information and Quantum Physics, University of Science and Technology of China, Hefei, Anhui 230026, China}
\affiliation{
\mbox{Hefei National Laboratory, University of Science and Technology of China, Hefei 230088, China}}

\date{\today}

\begin{abstract}
Zero- and ultralow-field nuclear magnetic resonance (ZULF NMR) has experienced rapid development and provides an excellent tool for diverse research fields ranging from materials science, quantum information processing to fundamental physics.
The detection of ZULF NMR signals in samples with natural abundance remains a challenging endeavor, due to the limited sensitivity of NMR detectors and thermal polarization.
In this work, we demonstrate a femtotesla potassium spin-exchange relaxation-free (SERF) magnetometer designed for ZULF NMR detection.
A potassium vapor cell with high buffer gas pressure and high atomic number density is used in the magnetometer.
With absorption spectroscopy and SERF effect,
the key parameters of the vapor cell are characterized and applied to optimize the magnetometer sensitivity.
To combine our SERF magnetometer and ZULF NMR detection, a custom-made vacuum chamber is employed to keep NMR sample close to the magnetometer cell and protect the sample from undesired heating effects.
Gradiometric measurement is performed to greatly reduce the magnetic noise.
With the phase calibration applied, the gradiometric measurement achieves 7-fold enhancement in magnetic-field sensitivity compared to the single channel and has a magnetic noise floor of 1.2\,fT/Hz$^{1/2}$.
Our SERF magnetometer exhibits high sensitivity and is promising to realize ZULF NMR detection of samples with natural abundance.
\end{abstract}
\maketitle

Over the past few decades, significant advancements have occurred in zero- and ultralow-field nuclear magnetic resonance (ZULF NMR)\,\cite{BLUMICH2009231,PhysRevLett.107.107601,doi:https://doi.org/10.1002/9780470034590.emrstm1369,10.1063/1.5003347,JIANG202168},
establishing it as a prominent research tool across various disciplines, including materials science\,\cite{PhysRevLett.107.107601,THEIS2013160,PhysRevB.92.220202}, quantum information processing\,\cite{doi:10.1126/sciadv.aar6327,ji2018time,jiang2018numerical,bian2017universal,tayler2016nuclear}, and tests of fundamental physics\,\cite{Garcon_2018,doi:10.1126/sciadv.aax4539,PhysRevLett.122.191302,PhysRevLett.121.023202}. Facilitated by the development of quantum magnetometers as NMR detectors, the field of ZULF NMR has yielded substantial progresses, including Superconducting Quantum Interference Devices (SQUIDs)\,\cite{RevModPhys.70.175,doi:10.1126/science.1069280,10.1063/1.2006981,10.1063/1.4976823} and spin-exchange relaxation-free (SERF)\,\cite{PhysRevLett.89.130801,Kominis2003,ledbetter2008spin,PhysRevApplied.11.024005,li2018serf} magnetometers. In contrast to high-field NMR, ZULF NMR achieves a high level of absolute field inhomogeneity, resulting in very narrow resonance lines\,\cite{PhysRevLett.107.107601,PhysRevLett.112.077601,https://doi.org/10.1002/qute.202000078,doi:https://doi.org/10.1002/9780470034590.emrstm1369,10.1063/1.5003347}.
Consequently, the ZULF NMR offers a robust technique to realize high-resolution spectroscopy for materials science, enabling tasks like chemical analysis utilizing \textit{J}-coupling multiplets\,\cite{THEIS2013160,PhysRevB.92.220202,PhysRevA.81.023420} and the determination of organic compound structures\,\cite{Blanchard2013,Alcicek2021}. Furthermore, recent investigations have unveiled the interaction between axionlike particles and nuclear spins, leading to the creation of oscillating pseudomagnetic fields that exert an influence on the nuclear spins. Consequently, ZULF NMR shows promising applications in the exploration of quantum-gravitational effects and the identification of axionlike dark matter\,\cite{Garcon_2018,doi:10.1126/sciadv.aax4539,BUDKER201966,PhysRevLett.122.191302,PhysRevLett.121.023202}.

\begin{figure}[b]  
	\makeatletter
	\def\@captype{figure}
	\makeatother
	\includegraphics[scale=0.60]{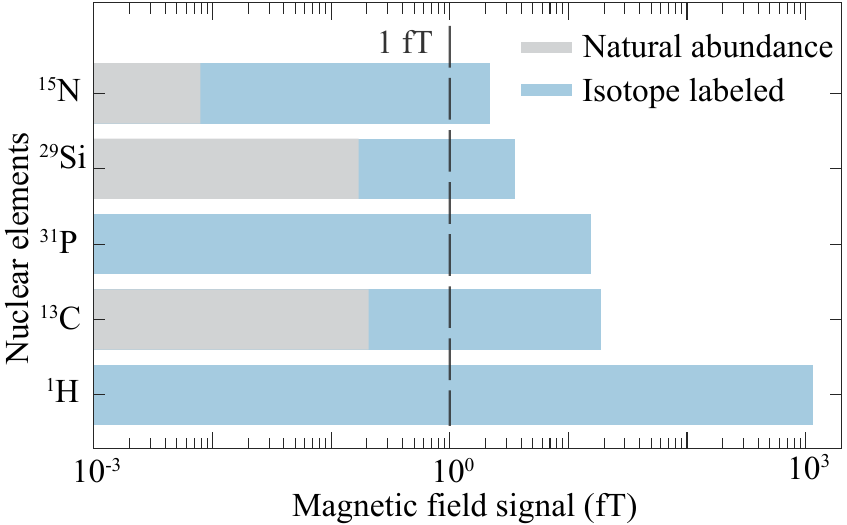} %
	\caption{Estimated magnetic field strength of NMR samples. The magnetic field signals are calculated under the condition that 200\,$\mu$L NMR samples are prepolarized by a field of 2\,T and are 1\,cm from the NMR detector. Areas shaded in gray show the magnetic field signals generated by the samples of natural abundance. Areas shaded by blue show the magnetic field signals of the isotope labeled samples. The vertical dash shows a magnetic field of 1\,fT.}
	\label{fig0}
\end{figure}

\begin{figure*}[t]  
	\makeatletter
	\def\@captype{figure}
	\makeatother
	\includegraphics[scale=1.6]{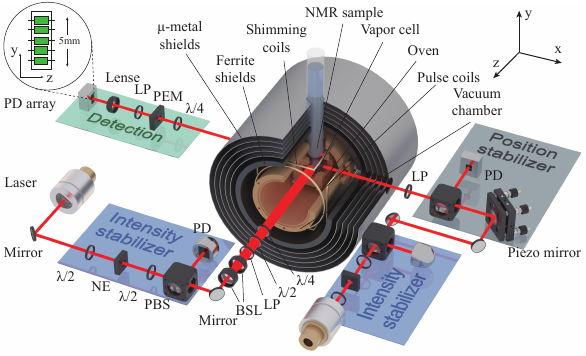} %
	\caption{Experimental setup of the ZULF NMR spectrometer. The intensity stabilizer mainly compromises of a NE, a PBS and a PD. The laser beam passes through the NE, and a small portion split by the PBS was detected by the PD. The intensity detected is fed back to the NE, and control the voltage of NE to stabilize the output laser intensity. The position stabilizer mainly compromises of a piezo mirror, a PBS, and a position detector. The circularly polarized pump beam, with an optical power of 100\,mW, transmits along the $z$-direction and is tuned at D1 line of K. The probe beam, with an optical power of 10\,mW, transmits along the $x$-direction, and is detected with two adjacent channels of the PD array. $\lambda$/2: half wave plate, NE: noise eater, PBS: polarizing beam splitter, PD: photodiode, BSL: beam shaping lenses, LP: linear polarizer, $\lambda$/4: quarter wave plate, PEM: photo-elastic modulator.    }
	\label{fig1}
\end{figure*}

Although great progresses have been achieved in ZULF NMR, there are still many challenges, in particular the detection of ZULF NMR in samples with natural abundance. As illustrated in Fig.\,\ref{fig0}, the magnetic field strengths produced by different nuclear elements with isotope labeled range from 1-1000\,fT.
However, the magnetic field strengths of NMR samples with natural abundance are below 1\,fT, posing a significant detection challenge for existing magnetometers that are used in ZULF NMR.
Increasing the polarization of NMR sample is an important approach to enhance magnetic-field signals.
Hyperpolarization, such as parahydrogen-induced polarization (PHIP), can produce strong signal of natural-abundance samples for detection in the ZULF regime\,\cite{Theis2011,doi:10.1126/sciadv.abp9242,Put2023}.
However, PHIP is only suitable for specific NMR samples,
in contrast,
the use of strong permanent magnets to thermally polarize samples is a general approach in NMR detection. In case of thermal polarization, the improvement of magnetometer sensitivity is a dominant issue.
As the most sensitive magnetometer currently, SERF magnetometer provides an excellent approach for ZULF NMR detection\,\cite{PhysRevLett.107.107601,doi:https://doi.org/10.1002/9780470034590.emrstm1369,10.1063/1.5003347,JIANG202168}. The sensitivity of state-of-the-art SERF magnetometers is primarily limited by magnetic-field noise due to Johnson noise of the $\mu$-metal shields\,\cite{Kominis2003,10.1063/1.3491215}. Employing ferrite shield to reduce the Johnson noise, the SERF magnetometer demonstrated a subfemtotesla-level sensitivity\,\cite{10.1063/1.3491215}.
However, it is still difficult to detect ZULF NMR signal with this kind of magnetometer. The vapor cell in this magnetometer was heated to high temperature around 473\,K and unsuitable for the detection of ZULF NMR, because leakage of thermal radiation can heat the NMR sample and consequently cause shift of nuclear spin-spin \textit{J} coupling and spectral line broadening\,\cite{doi:10.1021/j100859a001}. In addition, the volatilization of heated sample causes decrease in NMR signal, due to the signal strength is proportional to number of particles.

In this work, we present our experimental results of a femtotesla SERF magnetometer designed for ZULF NMR detection in samples with natural abundance. Compared with previous magnetometers for ZULF NMR, many technological innovations are made on this platform to improve the magnetometer sensitivity. A potassium vapor cell with high buffer gas pressure and high atomic number density is used in the magnetometer. By absorption spectroscopy and SERF effect, we estimate the key parameters of the vapor cell that are important in the optimization of magnetometer. To combine our SERF magnetometer and ZULF NMR detection, a custom-made vacuum chamber is employed to keep NMR sample close to the sensor and protect the sample from being heated. We find and measure the phase difference between two channels by applying oscillating magnetic field with different frequencies. We demonstrate that with proper phase calibration, the gradiometric measurement achieves 7-fold enhancement in magnetic field sensitivity compared to the single channel. Our SERF magnetometer has a magnetic noise floor of 1.2\,fT/Hz$^{1/2}$ from 20\,Hz to 30\,Hz. We also discuss promising application of the comprehensive setup in our work, and the future optimization of our ZULF NMR spectrometer.

\begin{figure*}[t]  
	\makeatletter
	\def\@captype{figure}
	\makeatother
	\includegraphics[scale=1.28]{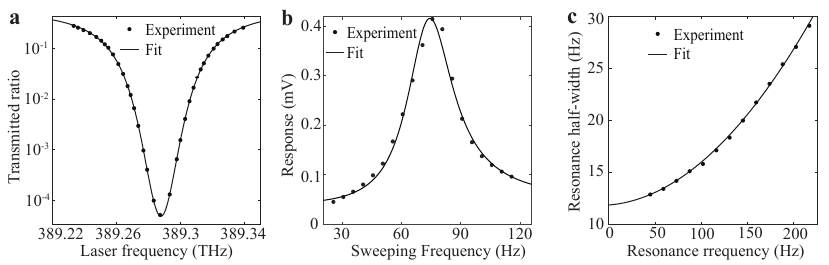} %
	\caption{The dynamics parameters of the vapor cell. (a) The absorption  spectrum which is measured at the D1 line of K and fitted to a Lorentzian, which gives a center frequency and half width at half maximum of 389.2879\,THz and 31.98\,GHz. (b) The frequency response fitted to a response curve. The fit gives a resonance linewidth of 10.45\,Hz. (c) Spin-exchange broadening of the resonance linewidth fitted by Eq.\,\ref{eqn1}. The fit gives $T_{\mathrm{SE}} \approx$  8.6\,$\mu$s. }
	\label{fig2}
\end{figure*}

Figure\,\ref{fig1} shows the apparatus of our ZULF NMR. A potassium atomic magnetometer is used to detect the magnetic field produced by NMR sample. The used vapor cell is spherical with a diameter of 2\,cm, and contains quenching gas $\mathrm{N}_2$, buffer gas $^{4}\mathrm{He}$, and a drop of potassium. 
The wall of the vapor cell is made of aluminosilicate glass (GE180 glass) to prevent the leakage of buffer gas $^{4}\mathrm{He}$.
The vapor cell is heated to 473\,K in a cubic boron-nitride oven by resistive heating with twisted coils carrying an alternating current of 125\,kHz. To reduce temperature gradients within the vapor cell, the boron-nitride oven is heated by six sets of twisted coils. The temperature of the vapor cell is measured with a thermistor (Pt\,1000) and fed to a Proportion Integration Differentiation (PID) controller, which ensures precise temperature control and maintains the vapor cell temperature within a 5\,mK deviation from the set value.
The oven is placed in a vacuum chamber (Fig.\,\ref{fig1}), and the vacuum chamber operates at approximately 10\,Pa. The vacuum chamber can reduce the heat leakage of the oven and protect the surrounding magnetic shields from being heated by the thermal radiation of the oven, which can cause an increase in Johnson noise generated by the magnetic shields. In addition, the vacuum chamber keeps the probe beam from the interference of hot air flow and reduces the noise at low frequency.
The vacuum chamber is placed within magnetic shields consisting of five-layer cylindrical $\mu$-metal and a layer of cylindrical Mn-Zn ferrite. The NMR tube containing sample can be transferred into ZULF detection region and about 1\,mm above the K vapor cell.

Low field and high atomic number density are achieved and thus the magnetometer operates in SERF regime\,\cite{PhysRevLett.89.130801,Kominis2003,ledbetter2008spin}, where the rate of spin-exchange collisions between K atoms greatly exceeds the Larmor precession of K spins. The pump beam and probe beam are generated by DBR lasers and propagate along the $z$- and $x$-direction, respectively. The pump beam is expanded to have a waist radius of about 1\,cm to cover the vapor cell. The power of pump beam is amplified to 100\,mW by a tapered amplifier. The circularly polarized pump beam is tuned close to the D1 transition of potassium at 770.1\,nm, and polarizes the K atoms along the $z$-direction. The magnetic field is measured via optical rotation of a 10-mW linearly polarized probe beam tuned at the D2 transition of K atoms. To reduce the noise caused by the laser power fluctuation, the pump beam and probe beam are stabilized with the intensity stabilizer (Fig.\,\ref{fig1}), which compromises of a noise eater, a polarizing beam splitter and a photodiode (PD). Furthermore, a position stabilizer is used to reduce the noise caused by the vibration of the laser beam. The transmitted probe beam is modulated by a photo-elastic modulator (PEM) to 50\,kHz, which can separate the signal from the low-frequency noise. The probe beam is detected with two photodiodes (top and bottom channels along the $y$-direction) in the PD array, where each photodiode is 1\,mm from the adjacent one. The output signal is then sent to a lock-in amplifier for demodulation,
and then the demodulated signal is recorded with a data acquisition device.

In order to optimize the magnetic sensitivity of our magnetometer, it is important to characterize the parameters of vapor cell, including buffer-gas pressure and atomic number density of potassium. High-pressure buffer gas can reduce the diffusion rate of potassium and prolong the relaxation time. The absorption spectrum is measured to estimate the pressure of $^4$He buffer gas in the vapor cell. The laser frequency is set at the D1 line of potassium and frequency sweeping is performed from 389.24\,THz to 389.34\,THz. The power of the incident and transmitted beam are measured and the ratio is a function of laser frequency. 
Figure\,\ref{fig2}a shows the absorption spectrum of the vapor cell. The experimental data is fitted to a Lorentzian profile, which gives out the center frequency and half width at half maximum of 389.2879\,THz and 31.98\,GHz. The difference of center frequency from the D1 line and the broadening of linewidth is due to the presence of buffer gas and quenching gas. For K atoms, the pressure shifts of D1 line are 3.9\,GHz/amg for $^4$He and -15.7\,GHz/amg for N$_2$. The pressure broadening widths are 13.3\,GHz/amg for $^4$He and 21.0\,GHz/amg for N$_2$\,\cite{Seltzer}. As a result, the contents of the vapor cell is calculated to be 1.86\,amg $^4$He and 0.34\,amg N$_2$.

\begin{figure}[t]  
	\makeatletter
	\def\@captype{figure}
	\makeatother
	\includegraphics[scale=0.7]{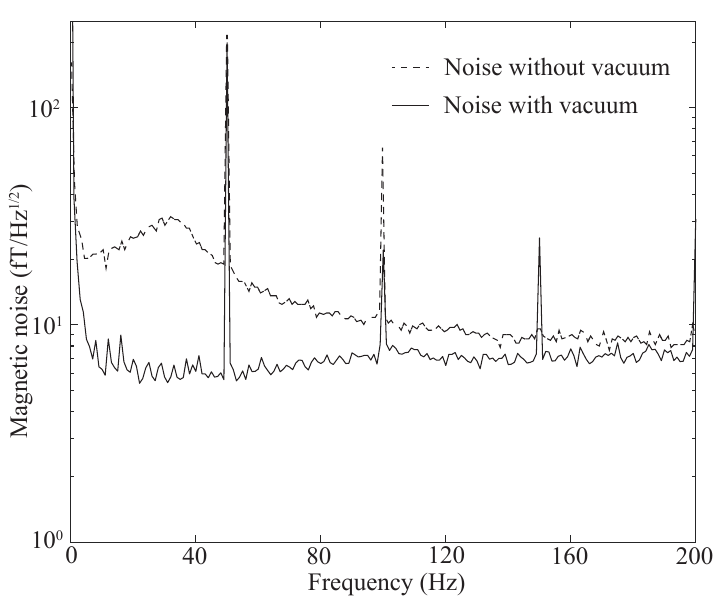} %
	\caption{ Magnetic-field noise of the SERF magnetometer. The noise of SERF magnetometer without vacuum chamber is has noise floor of 10\,fT/Hz$^{1/2}$ (dash line). The noise of SERF magnetometer with vacuum chamber is has noise floor of 8\,fT/Hz$^{1/2}$ (solid line).    }
	\label{fig-single}
\end{figure}

The traditional method to obtain the atomic number density of potassium is calculating with the absorption spectrum. However, our magnetometer operates at high temperature and absorption spectrum can not measured due to significant absorption of the laser beam around resonance frequency.
In order to solve this problem,
the method of measuring SERF effect is employed to estimate the atomic number density of potassium.
We explain the details as follows.
In small magnetic field and low atomic polarization, K spin exchange contributes to transverse relaxation $T_2^{\mathrm{SE}}$ only in second order and vanishes for zero field\,\cite{PhysRevLett.89.130801},         
\begin{equation}
    \frac{1}{T_2^{\mathrm{SE}}}=\omega_0^2T_{\mathrm{SE}}\left[\frac{1}{2}-\frac{(2I+1)^2}{2q^2}\right]q^2,
    \label{eqn1}
\end{equation}
where $\omega_0$ is K spin resonance frequency, \textit{I} is the nuclear spin, and \textit{q} is the slowing-down factor due to nuclear angular momentum. 
Different bias magnetic fields \textbf{B} are applied along the $z$-direction with the shimming coils (see Fig.\,\ref{fig1}). The bias magnetic field is smaller than 50\,nT and the intensity of pump beam is 0.2\,mW/cm$^2$ to keep the atoms in small magnetic field and weak polarization. The range of sweeping frequency includes resonance frequency where the response reach the peak. The frequency response is shown in Fig.\,\ref{fig2}b, and fitted to the response function. The resonance frequency and linewidth obtained by sweeping frequency are shown in Fig.\,\ref{fig2}c, and fitted by Eq.\,\ref{eqn1}.
The atomic number density $n$ is determined by the fitting parameter $T_{\mathrm{SE}}$ according to $1/T_{\mathrm{SE}} = n \Bar{v}\sigma_{\mathrm{SE}}$, where $\Bar{v} \approx$ 500\,m/s is the relative thermal velocity and \mbox{$\sigma_{\mathrm{SE}} \approx$ 2\,$\times$\,10$^{-14}$\,cm$^2$} is the spin-exchange cross section\,\cite{article}.
The fitting result gives $T_{\mathrm{SE}} \approx$ 8.6\,$\mu$s and the corresponding atomic number density is $n\approx$1.2\,$\times$\,10$^{14}$\,cm$^{-3}$.
The obtained high atomic number density can provide the magnetometer with a strong response to the magnetic signal.

\begin{figure}[t]  
	\makeatletter
	\def\@captype{figure}
	\makeatother
	\includegraphics[scale=0.7]{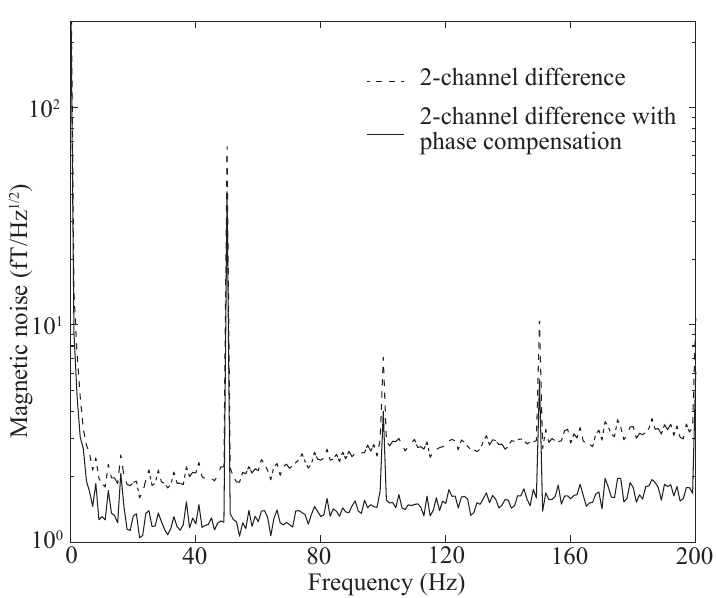} %
	\caption{Gradient magnetic-field noise of the SERF magnetometer. The subtraction of the two channels with amplitude calibration has a noise floor of 2.5\,fT/Hz$^{1/2}$ (dash line). The subtraction of the two channels with phase and amplitude calibration has a noise floor of 1.2\,fT/Hz$^{1/2}$ (solid line).  }
	\label{fig-2channel}
\end{figure}

The magnetic noise of our magnetometer is measured in the SERF regime.
 To calibrate the sensitivity, an oscillating magnetic field $\boldsymbol{B}_y$ = B$_y$sin($\omega$t)$\hat{y}$ with B$_y \approx$ 16\,pT at the frequency of 10\,Hz is applied along the $y$-direction by the shimming coils. The magnetic-field noise is about 8\,fT/Hz$^{1/2}$ from 10\,Hz to 200\,Hz (the solid line in Fig.\,\ref{fig-single}). The noise is dominated by the Johnson noise from $\mu$-metal shields. As a contrast, the noise of the magnetometer without vacuum chamber is also measured and shown in Fig.\,\ref{fig-single} (dash line). The noise is about 10\,fT/Hz$^{1/2}$ from 100\,Hz to 200\,Hz, but has larger noise at low frequency (especially below 40\,Hz) because the laser beams are disturbed by the hot air flow. Improving the performance of SERF magnetometer at low frequency is important for ZULF NMR detection, due to the signals of many samples are at low frequency, such as $^{13}$C-acetone and trimethyl phosphate\,\cite{ledbetter2013zero,appelt2006chemical,wilzewski2017method}.

The magnetic noise in single channel is dominated by the Johnson nosie from $\mu$-metal shields\,\cite{nenonen1996thermal}. To improve the sensitivity of our magnetometer,
we construct an atomic magnetic gradiometer by taking the difference between two adjacent channels of the photodiode array, which cancels the common magnetic field noise. Before subtracting, amplitude calibration are performed on top and bottom channels. In amplitude calibration, the parameter is obtained from the ratio between responses of two channels to the oscillating calibration field $\boldsymbol{B}_y$.  The subtraction of top and bottom channels is below 3\,fT/Hz$^{1/2}$ from 10\,Hz to 200\,Hz and has a noise floor of about 2.5\,fT/Hz$^{1/2}$ from 10\,Hz to 40\,Hz (the dash line in Fig.\,\ref{fig-2channel}). Compared to single channel, the magnetic sensitivity is improved by a factor of 3.2.

\begin{figure}[t]  
	\makeatletter
	\def\@captype{figure}
	\makeatother
	\includegraphics[scale=0.7]{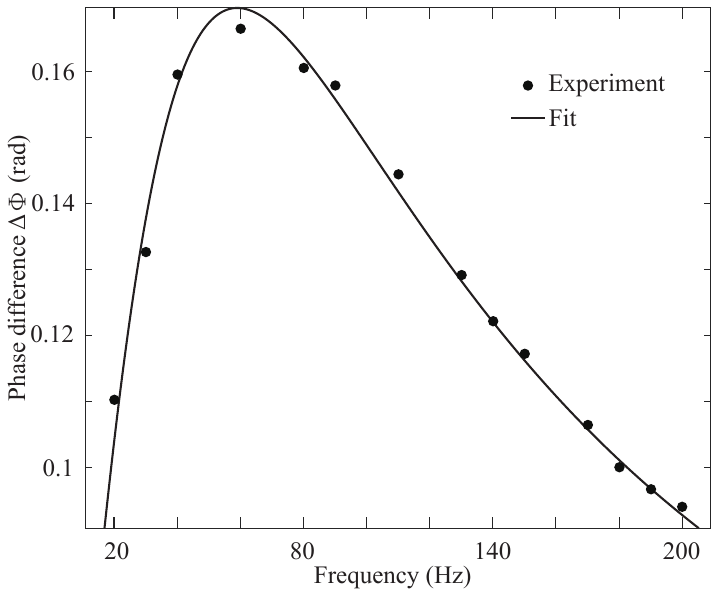} %
	\caption{The dependence of phase difference between top and bottom channels on the frequency. The experimental result is fitted to $\Delta \Phi =\mathrm{arctan}[f(f_1-f_2)/(f^2+f_1  f_2)]$, where the $f_1 \approx$  49.9\,Hz and $f_2 \approx$  68.8\,Hz are the bandwidth of the two channels.}
	\label{fig-phase}
\end{figure}

We show that there exists phase difference of response to signal between two channels,
where the phase difference has a detrimental effect on the gradiometric measurements. The phase difference $\Delta \Phi$ is deduced with the different bandwidths of two channels and is a function of frequency $f$,
 \begin{equation}
     \Delta \Phi = \mathrm{arctan}[f(f_1-f_2)/(f^2+f_1  f_2)],
     \label{eqn2}
 \end{equation} 
where $f_1$ and $f_2$ are the bandwidths of two channels.
The phase difference between two channels is measured by applying an oscillating magnetic field with different frequencies. The field is generated by a function generator (DS345, Standford Research Systems) and the shimming coils. The amplitude and phase of applied field are same for two channels. The experimental result is shown in Fig.\,\ref{fig-phase} and fitted to Eq.\,\ref{eqn2}. The fitting result gives $f_1 \approx$  49.9\,Hz and $f_2 \approx$  68.8\,Hz, which are the bandwidths of two channels. The phase difference reaches the maximum around 0.17\,rad when the frequency is \mbox{$f = \sqrt{f_1  f_2} \approx$ 58.6\,Hz}, which agrees well with the experimental data.

Employing proper phase calibration, we show that the common magnetic field noise can be further reduced. 
To calibrate the phase difference at all frequencies, the subtraction is performed after the conversion to frequency domain. The procedure involves phase-shifting the bottom complex Lorentzian by the angle obtained in Fig.\,\ref{fig-phase} while keeping the top channel unchanged. As depicted in Fig.\,\ref{fig-2channel}, the magnetic noise is below 1.7\,fT/Hz$^{1/2}$ from 10\,Hz to 200\,Hz, and is approximately 1.2\,fT/Hz$^{1/2}$ from 20\,Hz to 30\,Hz. With the amplitude and phase calibration, the difference of two channels achieves 7-fold improvement in magnetic-field sensitivity compared to the single channel.
To analyze the magnetic noise, the maximum reduction ratio is measured with an oscillating field of 20\,Hz applied. Comparing the signals in single channel and 2-channel difference, the reduction ratio is about 52.3. The reduction ratio of magnetic noise is limited due to the presence of gradient magnetic noise. 
With proper phase calibration, it is possible to eliminate the common magnetic-field noise. However, this could not suppress the gradient magnetic noise to the same degree. The gradient magnetic noise is mainly from the magnetic shields and the further optimization of magnetic shields could reduce the gradient magnetic noise.  


Based on the design of our femtotesla SERF magnetometer, it is feasible to detect the ZULF NMR signals of many nuclear elements. We propose to perform the ZULF NMR experiment, which is similar to many existing works\,\cite{blanchard2020zero,theis2013chemical,https://doi.org/10.1002/qute.202000078,doi:10.1126/sciadv.aar6327}. NMR sample (about 200\,$\mu$L) is contained in a standard 5-mm NMR tube, prepolarized in a permanent Halbach magnet ($B_p \approx$ 2.0\,T), and then pneumatically shuttled into the ultralow magnetic field detection region.
A guiding magnetic field ($\sim 10^{-4}$\,T) is applied during the transfer and is turned off quickly. Such a suddenly switched-off magnetic field serves effectively as
a radio-frequency pulse. After shuttled into the ultralow magnetic field, the initial state of the sample at zero field remains the high-field equilibrium state and evolves automatically under the Hamiltonian, which generates an oscillating magnetization signal and is measured by the atomic magnetometer.   

We would like to emphasize the main advantages of our magnetometer for ZULF NMR. The modules of intensity stabilizer and position stabilizer (shown in Fig.\,\ref{fig1}) are adopted to reduce the drift of laser power and position respectively, such that long-term detection is feasible to increase the signal-to-noise ratio (SNR) of NMR sample signal. During detection, the vapor cell is heated to a high temperature, but the NMR sample should be keep at room temperature. The vacuum chamber between them can protect the sample from being heated by the thermal radiation of vapor cell and keep the temperature of sample at about 310\,K. Furthermore, the photodiode array is used to perform gradiometric measurement of NMR signal, which can eliminate the common-mode noise and increase the SNR of NMR signal.

Our NMR spectrometer is in a very early stage of development, and many optimizations are in progress. To increase the SNR of NMR sample, a 2-Tesla permanent magnet is in construction which can provide higher prepolarization. Experimental implementation of universal quantum control in ZULF NMR is reported\,\cite{doi:10.1126/sciadv.aar6327}, which has a promissing application in the experiments to measure chiral molecules\,\cite{doi:10.1021/acs.jpclett.6b02653}. In addition, the interference effect in atomic magnetometers significantly improves the accuracy for determining the structure and molar concentration of NMR samples\,\cite{https://doi.org/10.1002/qute.202000078}.
Parahydrogen-induced polarization (PHIP) method is in preparation to provide higher polarization than permagnet\,\cite{Theis2011,theis2012zero,barskiy2019zero,natterer1997parahydrogen}.
In ZULF NMR, sample is typically thermally prepolarized in a permanent magnet, which suffers from low thermal polarization at the ambient temperature. To avoid this difficulty, hyperpolarization presents an alternative to the brute-force approach, and techniques such as PHIP are shown to produce signal for detection in the ZULF regime.

In conclusion, we demonstrate an femtotesla atomic magnetometer for ZULF NMR with a magnetic-field-gradient noise of 1.2\,fT/Hz$^{1/2}$. With gradiometric measurement, this SERF magnetometer is capable of detecting ZULF NMR signals in samples with natural isotopic abundance.
The buffer-gas pressure of the vapor cell is one of key factors in determining the performance of a SERF magnetometer. The spin-relaxation rate is a function of buffer gas pressure in the vapor cell. Optimal buffer gas pressure can be calculated to increase the response of magnetometer\,\cite{LU2022113928}. To reduce the gradient magnetic field noise, it is efficient to improve the shielding effect of the Mn-Zn ferrite\,\cite{10.1063/1.2737357}. The gradient noise can be reduced by enlarging the size of the Mn-Zn ferrite. The demagnetization of Mn-Zn ferrite can further reduce magnetic field noise including the gradient noise. The complete demagnetization of Mn-Zn ferrite require the use of circumferential and axial demagnetization coils. The demagnetization current driven in the coils should be very large due to the strong coercivity of Mn-Zn ferrite.  
 

\begin{acknowledgements}
    
This work was supported by the Innovation Program for Quantum Science and Technology (Grant No.\,2021ZD0303205),
National Natural Science Foundation of China (grants Nos.\,11661161018, 11927811, 12004371, 12150014, 12205296, 12274395), Youth Innovation Promotion Association (Grant No.\,2023474), National Key Research and Development Program of China (grant No.\,2018YFA0306600) and the Research Grants Council of Hong Kong (Grant No.\,14307420, 14308019,14309022).   
\end{acknowledgements}

\section*{data availability}

The data that support the findings of this study are available from the corresponding author upon reasonable request.

\bibliography{ref.bib}

\begin{thebibliography}{10}
\expandafter\ifx\csname url\endcsname\relax
  \def\url#1{\texttt{#1}}\fi
\expandafter\ifx\csname urlprefix\endcsname\relax\def\urlprefix{URL }\fi
\providecommand{\bibinfo}[2]{#2}
\providecommand{\eprint}[2][]{\url{#2}}

\bibitem{BLUMICH2009231}
\bibinfo{author}{Blümich, B.}, \bibinfo{author}{Casanova, F.} \&
  \bibinfo{author}{Appelt, S.}
\newblock \bibinfo{title}{Nmr at low magnetic fields}.
\newblock \emph{\bibinfo{journal}{Chemical Physics Letters}}
  \textbf{\bibinfo{volume}{477}}, \bibinfo{pages}{231--240}
  (\bibinfo{year}{2009}).
\newblock
  \urlprefix\url{https://www.sciencedirect.com/science/article/pii/S0009261409008070}.

\bibitem{PhysRevLett.107.107601}
\bibinfo{author}{Ledbetter, M.~P.} \emph{et~al.}
\newblock \bibinfo{title}{Near-zero-field nuclear magnetic resonance}.
\newblock \emph{\bibinfo{journal}{Phys. Rev. Lett.}}
  \textbf{\bibinfo{volume}{107}}, \bibinfo{pages}{107601}
  (\bibinfo{year}{2011}).
\newblock
  \urlprefix\url{https://link.aps.org/doi/10.1103/PhysRevLett.107.107601}.

\bibitem{JIANG202168}
\bibinfo{author}{Jiang, M.} \emph{et~al.}
\newblock \bibinfo{title}{Zero- to ultralow-field nuclear magnetic resonance
  and its applications}.
\newblock \emph{\bibinfo{journal}{Fundamental Research}}
  \textbf{\bibinfo{volume}{1}}, \bibinfo{pages}{68--84} (\bibinfo{year}{2021}).
\newblock
  \urlprefix\url{https://www.sciencedirect.com/science/article/pii/S2667325820300157}.

\bibitem{THEIS2013160}
\bibinfo{author}{Theis, T.} \emph{et~al.}
\newblock \bibinfo{title}{Chemical analysis using j-coupling multiplets in
  zero-field nmr}.
\newblock \emph{\bibinfo{journal}{Chemical Physics Letters}}
  \textbf{\bibinfo{volume}{580}}, \bibinfo{pages}{160--165}
  (\bibinfo{year}{2013}).
\newblock
  \urlprefix\url{https://www.sciencedirect.com/science/article/pii/S0009261413008191}.

\bibitem{RevModPhys.70.175}
\bibinfo{author}{Greenberg, Y.~S.}
\newblock \bibinfo{title}{Application of superconducting quantum interference
  devices to nuclear magnetic resonance}.
\newblock \emph{\bibinfo{journal}{Rev. Mod. Phys.}}
  \textbf{\bibinfo{volume}{70}}, \bibinfo{pages}{175--222}
  (\bibinfo{year}{1998}).
\newblock \urlprefix\url{https://link.aps.org/doi/10.1103/RevModPhys.70.175}.

\bibitem{doi:10.1126/science.1069280}
\bibinfo{author}{McDermott, R.} \emph{et~al.}
\newblock \bibinfo{title}{Liquid-state nmr and scalar couplings in microtesla
  magnetic fields}.
\newblock \emph{\bibinfo{journal}{Science}} \textbf{\bibinfo{volume}{295}},
  \bibinfo{pages}{2247--2249} (\bibinfo{year}{2002}).

\bibitem{10.1063/1.2006981}
\bibinfo{author}{Burghoff, M.}, \bibinfo{author}{Hartwig, S.},
  \bibinfo{author}{Trahms, L.} \& \bibinfo{author}{Bernarding, J.}
\newblock \bibinfo{title}{{Nuclear magnetic resonance in the nanoTesla range}}.
\newblock \emph{\bibinfo{journal}{Applied Physics Letters}}
  \textbf{\bibinfo{volume}{87}}, \bibinfo{pages}{054103}
  (\bibinfo{year}{2005}).

\bibitem{10.1063/1.4976823}
\bibinfo{author}{Storm, J.-H.}, \bibinfo{author}{Hömmen, P.},
  \bibinfo{author}{Drung, D.} \& \bibinfo{author}{Körber, R.}
\newblock \bibinfo{title}{{An ultra-sensitive and wideband magnetometer based
  on a superconducting quantum interference device}}.
\newblock \emph{\bibinfo{journal}{Applied Physics Letters}}
  \textbf{\bibinfo{volume}{110}}, \bibinfo{pages}{072603}
  (\bibinfo{year}{2017}).

\bibitem{PhysRevLett.89.130801}
\bibinfo{author}{Allred, J.~C.}, \bibinfo{author}{Lyman, R.~N.},
  \bibinfo{author}{Kornack, T.~W.} \& \bibinfo{author}{Romalis, M.~V.}
\newblock \bibinfo{title}{High-sensitivity atomic magnetometer unaffected by
  spin-exchange relaxation}.
\newblock \emph{\bibinfo{journal}{Phys. Rev. Lett.}}
  \textbf{\bibinfo{volume}{89}}, \bibinfo{pages}{130801}
  (\bibinfo{year}{2002}).
\newblock
  \urlprefix\url{https://link.aps.org/doi/10.1103/PhysRevLett.89.130801}.

\bibitem{10.1063/1.3491215}
\bibinfo{author}{Dang, H.~B.}, \bibinfo{author}{Maloof, A.~C.} \&
  \bibinfo{author}{Romalis, M.~V.}
\newblock \bibinfo{title}{{Ultrahigh sensitivity magnetic field and
  magnetization measurements with an atomic magnetometer}}.
\newblock \emph{\bibinfo{journal}{Applied Physics Letters}}
  \textbf{\bibinfo{volume}{97}}, \bibinfo{pages}{151110}
  (\bibinfo{year}{2010}).

\bibitem{Kominis2003}
\bibinfo{author}{Kominis, I.~K.}, \bibinfo{author}{Kornack, T.~W.},
  \bibinfo{author}{Allred, J.~C.} \& \bibinfo{author}{Romalis, M.~V.}
\newblock \bibinfo{title}{A subfemtotesla multichannel atomic magnetometer}.
\newblock \emph{\bibinfo{journal}{Nature}} \textbf{\bibinfo{volume}{422}},
  \bibinfo{pages}{596--599} (\bibinfo{year}{2003}).
\newblock \urlprefix\url{https://doi.org/10.1038/nature01484}.

\bibitem{PhysRevB.92.220202}
\bibinfo{author}{Blanchard, J.~W.} \emph{et~al.}
\newblock \bibinfo{title}{Measurement of untruncated nuclear spin interactions
  via zero- to ultralow-field nuclear magnetic resonance}.
\newblock \emph{\bibinfo{journal}{Phys. Rev. B}} \textbf{\bibinfo{volume}{92}},
  \bibinfo{pages}{220202} (\bibinfo{year}{2015}).
\newblock \urlprefix\url{https://link.aps.org/doi/10.1103/PhysRevB.92.220202}.

\bibitem{Blanchard2013}
\bibinfo{author}{Blanchard, J.~W.} \emph{et~al.}
\newblock \bibinfo{title}{High-resolution zero-field nmr j-spectroscopy of
  aromatic compounds}.
\newblock \emph{\bibinfo{journal}{Journal of the American Chemical Society}}
  \textbf{\bibinfo{volume}{135}}, \bibinfo{pages}{3607--3612}
  (\bibinfo{year}{2013}).
\newblock \urlprefix\url{https://doi.org/10.1021/ja312239v}.

\bibitem{Alcicek2021}
\bibinfo{author}{Alcicek, S.}, \bibinfo{author}{Put, P.},
  \bibinfo{author}{Kontul, V.} \& \bibinfo{author}{Pustelny, S.}
\newblock \bibinfo{title}{Zero-field nmr j-spectroscopy of organophosphorus
  compounds}.
\newblock \emph{\bibinfo{journal}{The Journal of Physical Chemistry Letters}}
  \textbf{\bibinfo{volume}{12}}, \bibinfo{pages}{787--792}
  (\bibinfo{year}{2021}).
\newblock \urlprefix\url{https://doi.org/10.1021/acs.jpclett.0c03532}.

\bibitem{Garcon_2018}
\bibinfo{author}{Garcon, A.} \emph{et~al.}
\newblock \bibinfo{title}{The cosmic axion spin precession experiment (casper):
  a dark-matter search with nuclear magnetic resonance}.
\newblock \emph{\bibinfo{journal}{Quantum Science and Technology}}
  \textbf{\bibinfo{volume}{3}}, \bibinfo{pages}{014008} (\bibinfo{year}{2017}).
\newblock \urlprefix\url{https://dx.doi.org/10.1088/2058-9565/aa9861}.

\bibitem{PhysRevLett.121.023202}
\bibinfo{author}{Wu, T.}, \bibinfo{author}{Blanchard, J.~W.},
  \bibinfo{author}{Jackson~Kimball, D.~F.}, \bibinfo{author}{Jiang, M.} \&
  \bibinfo{author}{Budker, D.}
\newblock \bibinfo{title}{Nuclear-spin comagnetometer based on a liquid of
  identical molecules}.
\newblock \emph{\bibinfo{journal}{Phys. Rev. Lett.}}
  \textbf{\bibinfo{volume}{121}}, \bibinfo{pages}{023202}
  (\bibinfo{year}{2018}).
\newblock
  \urlprefix\url{https://link.aps.org/doi/10.1103/PhysRevLett.121.023202}.

\bibitem{doi:10.1126/sciadv.aax4539}
\bibinfo{author}{Garcon, A.} \emph{et~al.}
\newblock \bibinfo{title}{Constraints on bosonic dark matter from
  ultralow-field nuclear magnetic resonance}.
\newblock \emph{\bibinfo{journal}{Science Advances}}
  \textbf{\bibinfo{volume}{5}}, \bibinfo{pages}{eaax4539}
  (\bibinfo{year}{2019}).

\bibitem{BUDKER201966}
\bibinfo{author}{Budker, D.}
\newblock \bibinfo{title}{Extreme nuclear magnetic resonance: Zero field,
  single spins, dark matter…}.
\newblock \emph{\bibinfo{journal}{Journal of Magnetic Resonance}}
  \textbf{\bibinfo{volume}{306}}, \bibinfo{pages}{66--68}
  (\bibinfo{year}{2019}).
\newblock
  \urlprefix\url{https://www.sciencedirect.com/science/article/pii/S1090780719301351}.

\bibitem{PhysRevLett.122.191302}
\bibinfo{author}{Wu, T.} \emph{et~al.}
\newblock \bibinfo{title}{Search for axionlike dark matter with a liquid-state
  nuclear spin comagnetometer}.
\newblock \emph{\bibinfo{journal}{Phys. Rev. Lett.}}
  \textbf{\bibinfo{volume}{122}}, \bibinfo{pages}{191302}
  (\bibinfo{year}{2019}).
\newblock
  \urlprefix\url{https://link.aps.org/doi/10.1103/PhysRevLett.122.191302}.

\bibitem{8425967}
\bibinfo{author}{Li, J.} \emph{et~al.}
\newblock \bibinfo{title}{Serf atomic magnetometer–recent advances and
  applications: A review}.
\newblock \emph{\bibinfo{journal}{IEEE Sensors Journal}}
  \textbf{\bibinfo{volume}{18}}, \bibinfo{pages}{8198--8207}
  (\bibinfo{year}{2018}).

\bibitem{LU2022113928}
\bibinfo{author}{Lu, J.} \emph{et~al.}
\newblock \bibinfo{title}{Optimal buffer gas pressure in dual-beam
  spin-exchange relaxation-free magnetometers}.
\newblock \emph{\bibinfo{journal}{Sensors and Actuators A: Physical}}
  \textbf{\bibinfo{volume}{347}}, \bibinfo{pages}{113928}
  (\bibinfo{year}{2022}).
\newblock
  \urlprefix\url{https://www.sciencedirect.com/science/article/pii/S0924424722005635}.

\bibitem{Seltzer}
\bibinfo{author}{Seltzer, S.~J.}
\newblock \emph{\bibinfo{title}{Developments in alkali -metal atomic
  magnetometry}}.
\newblock Ph.D. thesis (\bibinfo{year}{2008}).
\newblock
  \urlprefix\url{https://www.proquest.com/dissertations-theses/developments-alkali-metal-atomic-magnetometry/docview/275671021/se-2}.

\bibitem{PhysRevApplied.11.024005}
\bibinfo{author}{Jiang, M.} \emph{et~al.}
\newblock \bibinfo{title}{Magnetic gradiometer for the detection of zero- to
  ultralow-field nuclear magnetic resonance}.
\newblock \emph{\bibinfo{journal}{Phys. Rev. Appl.}}
  \textbf{\bibinfo{volume}{11}}, \bibinfo{pages}{024005}
  (\bibinfo{year}{2019}).
\newblock
  \urlprefix\url{https://link.aps.org/doi/10.1103/PhysRevApplied.11.024005}.

\bibitem{Theis2011}
\bibinfo{author}{Theis, T.} \emph{et~al.}
\newblock \bibinfo{title}{Parahydrogen-enhanced zero-field nuclear magnetic
  resonance}.
\newblock \emph{\bibinfo{journal}{Nature Physics}}
  \textbf{\bibinfo{volume}{7}}, \bibinfo{pages}{571--575}
  (\bibinfo{year}{2011}).
\newblock \urlprefix\url{https://doi.org/10.1038/nphys1986}.

\bibitem{10.1063/1.2737357}
\bibinfo{author}{Kornack, T.~W.}, \bibinfo{author}{Smullin, S.~J.},
  \bibinfo{author}{Lee, S.-K.} \& \bibinfo{author}{Romalis, M.~V.}
\newblock \bibinfo{title}{{A low-noise ferrite magnetic shield}}.
\newblock \emph{\bibinfo{journal}{Applied Physics Letters}}
  \textbf{\bibinfo{volume}{90}}, \bibinfo{pages}{223501}
  (\bibinfo{year}{2007}).

\bibitem{doi:10.1126/sciadv.aar6327}
\bibinfo{author}{Jiang, M.} \emph{et~al.}
\newblock \bibinfo{title}{Experimental benchmarking of quantum control in
  zero-field nuclear magnetic resonance}.
\newblock \emph{\bibinfo{journal}{Science Advances}}
  \textbf{\bibinfo{volume}{4}}, \bibinfo{pages}{eaar6327}
  (\bibinfo{year}{2018}).

\bibitem{doi:10.1021/j100859a001}
\bibinfo{author}{Brey, W. S.~J.}, \bibinfo{author}{Scott, K.~N.} \&
  \bibinfo{author}{Whitman, D.~R.}
\newblock \bibinfo{title}{Temperature dependence of nuclear magnetic resonance
  coupling constants and chemical shifts of the vinyl halides and some vinyl
  esters}.
\newblock \emph{\bibinfo{journal}{The Journal of Physical Chemistry}}
  \textbf{\bibinfo{volume}{72}}, \bibinfo{pages}{4351--4359}
  (\bibinfo{year}{1968}).

\bibitem{https://doi.org/10.1002/qute.202000078}
\bibinfo{author}{Jiang, M.} \emph{et~al.}
\newblock \bibinfo{title}{Interference in atomic magnetometry}.
\newblock \emph{\bibinfo{journal}{Advanced Quantum Technologies}}
  \textbf{\bibinfo{volume}{3}}, \bibinfo{pages}{2000078}
  (\bibinfo{year}{2020}).
\newblock
  \urlprefix\url{https://onlinelibrary.wiley.com/doi/abs/10.1002/qute.202000078}.

\bibitem{PhysRevLett.112.077601}
\bibinfo{author}{Emondts, M.} \emph{et~al.}
\newblock \bibinfo{title}{Long-lived heteronuclear spin-singlet states in
  liquids at a zero magnetic field}.
\newblock \emph{\bibinfo{journal}{Phys. Rev. Lett.}}
  \textbf{\bibinfo{volume}{112}}, \bibinfo{pages}{077601}
  (\bibinfo{year}{2014}).
\newblock
  \urlprefix\url{https://link.aps.org/doi/10.1103/PhysRevLett.112.077601}.

\bibitem{doi:https://doi.org/10.1002/9780470034590.emrstm1369}
\bibinfo{author}{Blanchard, J.~W.} \& \bibinfo{author}{Budker, D.}
\newblock \bibinfo{title}{Zero- to ultralow-field nmr}
  \bibinfo{pages}{1395--1410} (\bibinfo{year}{2016}).
\newblock
  \urlprefix\url{https://onlinelibrary.wiley.com/doi/abs/10.1002/9780470034590.emrstm1369}.

\bibitem{10.1063/1.5003347}
\bibinfo{author}{Tayler, M. C.~D.} \emph{et~al.}
\newblock \bibinfo{title}{{Invited Review Article: Instrumentation for nuclear
  magnetic resonance in zero and ultralow magnetic field}}.
\newblock \emph{\bibinfo{journal}{Review of Scientific Instruments}}
  \textbf{\bibinfo{volume}{88}}, \bibinfo{pages}{091101}
  (\bibinfo{year}{2017}).

\bibitem{PhysRevA.81.023420}
\bibinfo{author}{Appelt, S.} \emph{et~al.}
\newblock \bibinfo{title}{Paths from weak to strong coupling in nmr}.
\newblock \emph{\bibinfo{journal}{Phys. Rev. A}} \textbf{\bibinfo{volume}{81}},
  \bibinfo{pages}{023420} (\bibinfo{year}{2010}).
\newblock \urlprefix\url{https://link.aps.org/doi/10.1103/PhysRevA.81.023420}.

\bibitem{doi:10.1126/sciadv.abp9242}
\bibinfo{author}{Dyke, E. T.~V.} \emph{et~al.}
\newblock \bibinfo{title}{Relayed hyperpolarization for zero-field nuclear
  magnetic resonance}.
\newblock \emph{\bibinfo{journal}{Science Advances}}
  \textbf{\bibinfo{volume}{8}}, \bibinfo{pages}{eabp9242}
  (\bibinfo{year}{2022}).

\bibitem{Put2023}
\bibinfo{author}{Put, P.} \emph{et~al.}
\newblock \bibinfo{title}{Detection of pyridine derivatives by sabre
  hyperpolarization at zero field}.
\newblock \emph{\bibinfo{journal}{Communications Chemistry}}
  \textbf{\bibinfo{volume}{6}}, \bibinfo{pages}{131} (\bibinfo{year}{2023}).
\newblock \urlprefix\url{https://doi.org/10.1038/s42004-023-00928-z}.

\bibitem{article}
\bibinfo{author}{Aleksandrov, E., Alexandrov}, \bibinfo{author}{Balabas, M.},
  \bibinfo{author}{Vershovskii, A.}, \bibinfo{author}{A.I, O.} \&
  \bibinfo{author}{N.n, Y.}
\newblock \bibinfo{title}{Spin-exchange broadening of magnetic-resonance line
  of potassium atoms}.
\newblock \emph{\bibinfo{journal}{Optics and Spectroscopy}}
  \textbf{\bibinfo{volume}{87}}, \bibinfo{pages}{359} (\bibinfo{year}{1999}).

\bibitem{nenonen1996thermal}
\bibinfo{author}{Nenonen, J.}, \bibinfo{author}{Montonen, J.} \&
  \bibinfo{author}{Katila, T.}
\newblock \bibinfo{title}{Thermal noise in biomagnetic measurements}.
\newblock \emph{\bibinfo{journal}{Review of scientific instruments}}
  \textbf{\bibinfo{volume}{67}}, \bibinfo{pages}{2397--2405}
  (\bibinfo{year}{1996}).

\bibitem{li2018serf}
\bibinfo{author}{Li, J.} \emph{et~al.}
\newblock \bibinfo{title}{Serf atomic magnetometer--recent advances and
  applications: A review}.
\newblock \emph{\bibinfo{journal}{IEEE Sensors Journal}}
  \textbf{\bibinfo{volume}{18}}, \bibinfo{pages}{8198--8207}
  (\bibinfo{year}{2018}).

\bibitem{ledbetter2013zero}
\bibinfo{author}{Ledbetter, M.~P.} \& \bibinfo{author}{Budker, D.}
\newblock \bibinfo{title}{Zero-field nuclear magnetic resonance}.
\newblock \emph{\bibinfo{journal}{Physics Today}}
  \textbf{\bibinfo{volume}{66}}, \bibinfo{pages}{44--49}
  (\bibinfo{year}{2013}).

\bibitem{appelt2006chemical}
\bibinfo{author}{Appelt, S.}, \bibinfo{author}{K{\"u}hn, H.},
  \bibinfo{author}{H{\"a}sing, F.~W.} \& \bibinfo{author}{Bl{\"u}mich, B.}
\newblock \bibinfo{title}{Chemical analysis by ultrahigh-resolution nuclear
  magnetic resonance in the earth’s magnetic field}.
\newblock \emph{\bibinfo{journal}{Nature Physics}}
  \textbf{\bibinfo{volume}{2}}, \bibinfo{pages}{105--109}
  (\bibinfo{year}{2006}).

\bibitem{wilzewski2017method}
\bibinfo{author}{Wilzewski, A.}, \bibinfo{author}{Afach, S.},
  \bibinfo{author}{Blanchard, J.~W.} \& \bibinfo{author}{Budker, D.}
\newblock \bibinfo{title}{A method for measurement of spin-spin couplings with
  sub-mhz precision using zero-to ultralow-field nuclear magnetic resonance}.
\newblock \emph{\bibinfo{journal}{Journal of Magnetic Resonance}}
  \textbf{\bibinfo{volume}{284}}, \bibinfo{pages}{66--72}
  (\bibinfo{year}{2017}).

\bibitem{blanchard2020zero}
\bibinfo{author}{Blanchard, J.~W.}, \bibinfo{author}{Wu, T.},
  \bibinfo{author}{Eills, J.}, \bibinfo{author}{Hu, Y.} \&
  \bibinfo{author}{Budker, D.}
\newblock \bibinfo{title}{Zero-to ultralow-field nuclear magnetic resonance
  j-spectroscopy with commercial atomic magnetometers}.
\newblock \emph{\bibinfo{journal}{Journal of Magnetic Resonance}}
  \textbf{\bibinfo{volume}{314}}, \bibinfo{pages}{106723}
  (\bibinfo{year}{2020}).

\bibitem{theis2013chemical}
\bibinfo{author}{Theis, T.} \emph{et~al.}
\newblock \bibinfo{title}{Chemical analysis using j-coupling multiplets in
  zero-field nmr}.
\newblock \emph{\bibinfo{journal}{Chemical Physics Letters}}
  \textbf{\bibinfo{volume}{580}}, \bibinfo{pages}{160--165}
  (\bibinfo{year}{2013}).

\bibitem{theis2012zero}
\bibinfo{author}{Theis, T.} \emph{et~al.}
\newblock \bibinfo{title}{Zero-field nmr enhanced by parahydrogen in reversible
  exchange}.
\newblock \emph{\bibinfo{journal}{Journal of the American Chemical Society}}
  \textbf{\bibinfo{volume}{134}}, \bibinfo{pages}{3987--3990}
  (\bibinfo{year}{2012}).

\bibitem{barskiy2019zero}
\bibinfo{author}{Barskiy, D.~A.} \emph{et~al.}
\newblock \bibinfo{title}{Zero-field nuclear magnetic resonance of chemically
  exchanging systems}.
\newblock \emph{\bibinfo{journal}{Nature communications}}
  \textbf{\bibinfo{volume}{10}}, \bibinfo{pages}{3002} (\bibinfo{year}{2019}).

\bibitem{natterer1997parahydrogen}
\bibinfo{author}{Natterer, J.} \& \bibinfo{author}{Bargon, J.}
\newblock \bibinfo{title}{Parahydrogen induced polarization}.
\newblock \emph{\bibinfo{journal}{Progress in Nuclear Magnetic Resonance
  Spectroscopy}} \textbf{\bibinfo{volume}{31}}, \bibinfo{pages}{293--315}
  (\bibinfo{year}{1997}).

\bibitem{ledbetter2008spin}
\bibinfo{author}{Ledbetter, M.}, \bibinfo{author}{Savukov, I.},
  \bibinfo{author}{Acosta, V.}, \bibinfo{author}{Budker, D.} \&
  \bibinfo{author}{Romalis, M.}
\newblock \bibinfo{title}{Spin-exchange-relaxation-free magnetometry with cs
  vapor}.
\newblock \emph{\bibinfo{journal}{Physical Review A}}
  \textbf{\bibinfo{volume}{77}}, \bibinfo{pages}{033408}
  (\bibinfo{year}{2008}).

\bibitem{doi:10.1021/acs.jpclett.6b02653}
\bibinfo{author}{King, J.~P.}, \bibinfo{author}{Sjolander, T.~F.} \&
  \bibinfo{author}{Blanchard, J.~W.}
\newblock \bibinfo{title}{Antisymmetric couplings enable direct observation of
  chirality in nuclear magnetic resonance spectroscopy}.
\newblock \emph{\bibinfo{journal}{The Journal of Physical Chemistry Letters}}
  \textbf{\bibinfo{volume}{8}}, \bibinfo{pages}{710--714}
  (\bibinfo{year}{2017}).

\bibitem{ji2018time}
\bibinfo{author}{Ji, Y.}, \bibinfo{author}{Bian, J.}, \bibinfo{author}{Jiang,
  M.}, \bibinfo{author}{D'Alessandro, D.} \& \bibinfo{author}{Peng, X.}
\newblock \bibinfo{title}{Time-optimal control of independent spin-1/2 systems
  under simultaneous control}.
\newblock \emph{\bibinfo{journal}{Physical Review A}}
  \textbf{\bibinfo{volume}{98}}, \bibinfo{pages}{062108}
  (\bibinfo{year}{2018}).

\bibitem{jiang2018numerical}
\bibinfo{author}{Jiang, M.} \emph{et~al.}
\newblock \bibinfo{title}{Numerical optimal control of spin systems at zero
  magnetic field}.
\newblock \emph{\bibinfo{journal}{Physical Review A}}
  \textbf{\bibinfo{volume}{97}}, \bibinfo{pages}{062118}
  (\bibinfo{year}{2018}).

\bibitem{bian2017universal}
\bibinfo{author}{Bian, J.} \emph{et~al.}
\newblock \bibinfo{title}{Universal quantum control in zero-field nuclear
  magnetic resonance}.
\newblock \emph{\bibinfo{journal}{Physical Review A}}
  \textbf{\bibinfo{volume}{95}}, \bibinfo{pages}{052342}
  (\bibinfo{year}{2017}).

\bibitem{tayler2016nuclear}
\bibinfo{author}{Tayler, M.~C.}, \bibinfo{author}{Sjolander, T.~F.},
  \bibinfo{author}{Pines, A.} \& \bibinfo{author}{Budker, D.}
\newblock \bibinfo{title}{Nuclear magnetic resonance at millitesla fields using
  a zero-field spectrometer}.
\newblock \emph{\bibinfo{journal}{Journal of Magnetic Resonance}}
  \textbf{\bibinfo{volume}{270}}, \bibinfo{pages}{35--39}
  (\bibinfo{year}{2016}).

\end{thebibliography}
\end{document}